\documentclass[prl, twocolumn, showpacs]{revtex4}
\usepackage{graphicx}
\usepackage{amsmath}

\newcommand{\ket}[1]{|#1\rangle}

\newcommand{\expt}[1]{\langle #1\rangle}

\begin{document}

\title{Metallic Nature of the Four-Dimensional Quantum Hall Edge}

\author{Y. D. Chong}
\author{R. B. Laughlin}
  \affiliation{Department of Physics, Stanford University, Stanford,
    California 94309}

\date{\today}

\begin{abstract}
We study the density response of the four-dimensional quantum Hall
fluid found by Zhang and Hu, which has been advanced as a model of
emergent relativity. We calculate the density-density correlation
function along the edge at half-filling, and show that it is similar
to the three-dimensional free electron gas. This indicates that the
edge of the four-dimensional quantum Hall fluid behaves like a metal.
\end{abstract}

\pacs{71.10.Ca, 73.20.Mf, 73.43.-f}

\maketitle

Zhang and Hu have recently discovered a beautiful four-dimensional
generalization of the quantum Hall state \cite{ZH, ZH2}. The relevance
of this model to nature is the potential for emergent relativity at
its edge. Were there such an effect, the magnetic length---a scale
implicit in all Landau-level structures---would form a natural
ultraviolet cutoff for the system, potentially producing a
quantum-mechanical description of the vacuum of space-time that is
internally consistent at all length scales, something that presently
does not exist. To this end, Zhang and Hu discussed a set of special
particle-hole pair excitations obeying the ``equations of motion'' of
massless relativistic bosons. However, it is quite misleading to
describe the boson dispersion relation in this way, because these
objects are not real particles unless something stabilizes them
against decay into their fermionic components \cite{Polchinski}. There
are two strong reasons to suspect that this is not the case. One is
that the model consists fundamentally of noninteracting fermions
filled up to a Fermi energy. Since the density of states at the Fermi
energy is nonzero, the low-temperature specific heat of the model is
linear in the temperature, and thus vastly larger than any
relativistic system could possibly have; a relativistic model in 3
spatial dimensions has a specific heat rising as $T^3$ if it is
massless, and exponentially activated if it has mass. The second point
is that the fermions flow towards potential minima in just the way
electrons in a metal do; this is, in fact, a key assumption of Zhang
and Hu's analysis. The key properties of the model thus indicate that
it is not a relativistic vacuum, but a new kind of metal.

\begin{figure}
	    \includegraphics[scale=0.95]{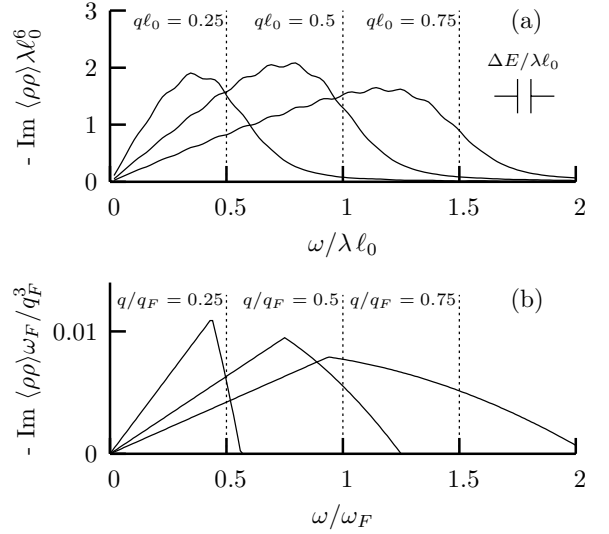}
      \caption{ Plots of $- {\rm Im} \expt{\rho_q \rho_{-q}}_\omega$
        against $\omega$. (a) The 4D quantum Hall edge (Eq.
        \eqref{fourier}), with $\eta = 0.087 \, \lambda \ell_0$,
        $\alpha = 1$, and $p = 300$. $q$ and $\omega$ are scaled
        against the natural units $1/\ell_0$ and $\lambda \ell_0$.
        Dotted lines are the threshold frequencies $\omega = 2
        (q\ell_0) (\lambda \ell_0)$. The wiggles in the curves are due
        to fact that the excitation energies are discrete (the energy
        quantum $\Delta E = [2 \sqrt{p}/(p+4)] \lambda\ell_0 $ is
        marked), which go to a continuum as $p \rightarrow \infty$.
        (b) The three-dimensional free electron gas (Eq.
        \eqref{Lindhard}), with $\omega$ and $q$ scaled against the
        Fermi frequency and wave vector respectively. Dotted lines
        show the frequencies $\omega = c q$, where $c =
        2\,\omega_F/q_F$ is the Fermi velocity. }
      \label{fig:pp vs q}
    \end{figure}

In order to clarify this matter, we have calculated the model's
density-density response function numerically. The result is shown in
Fig. \ref{fig:pp vs q}. It may be seen that this response function is
so much like that of a noninteracting Fermi sea that the two are {\it
  impossible to distinguish} at low frequencies and long wavelengths.
This suggests that the Zhang-Hu model is a metal, albeit one lacking a
conventional Fermi surface and therefore of a type not previously
known, and that the ``relativistic bosons'' identified by Zhang and Hu
are modes of compressional sound.

The Zhang-Hu model describes noninteracting fermions of isospin $p/2$
placed on the surface of a sphere in five spatial dimensions. Position
on the sphere is specified by coordinates $x = (x_1,.., x_5)$, which
one normalizes to the sphere radius: $\sum_{a=1}^5 x_a^2 = 1$. The
Hamiltonian is
\begin{equation}
{\cal H} = - \frac{1}{2} \sum_{a < b}
\left( x_a D_b - x_b D_a  \right)^2 + \lambda x_5
\end{equation}
where $D_a \equiv \partial_a + a_a$  and $\lambda x_5$ is a perturbative 
confining potential.  The vector potential $a_a$ is generated by a Yang 
monopole \cite{Yang},
\begin{align}
\begin{aligned}
a_\mu = \frac{-i}{2(1+x_5)} \eta_{\mu\nu}^i x_{\nu} I_i
\quad , \quad
a_5 = 0 \\
\eta_{\mu\nu}^i = \epsilon_{i\mu\nu4} + \delta_{i\mu} \delta_{4\nu}
- \delta_{i\nu} \delta_{4\mu}
\end{aligned}
\end{align}
where $I_i$ are SU(2) isospin operators satisfying $[I_i , I_j ] = i
\epsilon_{ijk} I_k$.  The  ``strength'' of the monopole is 
characterized by the eigenvalue of $I^2 = I_1^2 + I_2^2 +
I_3^2$, which is $p/2(p/2+1)$.

The natural length scale is the magnetic length, given by $\ell_0
\equiv 1/\sqrt{p}$. Zhang and Hu work in the thermodynamic limit $p
\rightarrow \infty$, meaning the magnetic length vanishes compared to
the sphere radius. The natural energy scale is $\lambda \ell_0$, the
confining potential energy at one magnetic length.

The isospin degree of freedom is conveniently parameterized by
coherent-state spinor coordinates $u_1 = \cos(\theta/2)e^{i\phi/2}$, $u_2
= \sin(\theta/2)e^{-i\phi/2}$ \cite{Haldane}. In this representation,
the isospin operators have the form
\begin{equation}
I_+ = u_1 \frac{\partial}{\partial u_2}
\; \;
I_- = u_2 \frac{\partial}{\partial u_1}
\; \; 
I_3 = \frac{1}{2} ( u_1 \frac{\partial}{\partial u_1} 
- u_2 \frac{\partial}{\partial u_2} ).
\end{equation}

In the absence of a confining potential ($\lambda = 0$), the energy
eigenstates of this Hamiltonian in the lowest Landau level are all
degenerate and given by
\begin{equation}
\expt{x, u|m} = \sqrt{\frac{p!}{m_1! m_2! m_3! m_4!}}
\Psi_1^{m_1} \Psi_2^{m_2} \Psi_3^{m_3} \Psi_4^{m_4}
\label{wavefunctions}
\end{equation}
where $m_1 + m_2 + m_3 + m_4 = p$ and
\begin{align}
\begin{aligned}
\genfrac{(}{)}{0pt}{}{\Psi_1}{\Psi_2} &= \sqrt{\frac{1+x_5}{2}}
\genfrac{(}{)}{0pt}{}{u_1}{u_2} \\
\genfrac{(}{)}{0pt}{}{\Psi_3}{\Psi_4} &=
\frac{x_4 - i x_i \sigma_i}{\sqrt{2(1+x_5)}}
\genfrac{(}{)}{0pt}{}{u_1}{u_2},
\end{aligned}
\end{align}
with $\sigma_i$ denoting the Pauli matrices. We abbreviate the set of four 
quantum numbers by $m = \{m_1, m_2, m_3, m_4\}$ and denote these states by
$\ket{m}$.

In the presence of a confining potential, these wavefunctions continue
to be eigenstates to linear order in $\lambda$, but the degeneracy is
partially broken because they are centered at different latitudes. The
energy eigenvalue of $\ket{m}$ is $E_m = \lambda \expt{m |x_5|m}$,
where
\begin{equation}
\expt{m|x_5|m} = \frac{m_1 + m_2 - m_3 - m_4}{p+4}
\; \sqrt{p} \; \ell_0 \; \; .
\end{equation}
If the number of particles $N$ is less than the Landau level degeneracy
$(p+1)(p+2)(p+3)/6$, the confining potential causes the particles to
puddle at the bottom of the sphere.  The multi-particle ground state is
then a Slater determinant of single-particle states having $\expt{x_5}$ up
to some ``Fermi latitude'' $x_5^F$. The edge of the puddle, given by $x_5
= x_5^F$, is a three-dimensional space. With $N$ fixed, multi-particle
excited states are constructed by introducing particle-hole pairs, with
hole states centered below the Fermi latitude and particle states above
it. We ignore excitations involving states in higher Landau levels, which
is permissible since the confining potential is weak.

The density-density response function \cite{Pines} describes the
changes to the particle density that result from perturbing the system
with weak potentials, and is effectively the Ohm's law conductivity.
Analogies to the conventional Ohm's law must be drawn carefully here
because the appropriate conserved particle current is non-Abelian. The
density-density response function describes only the longitudinal
conductivity measured, for example, in an electron energy-loss
experiment in aluminium film. This kind of experiment also makes sense
in the Zhang-Hu model because the number of fermions is conserved. The
idea is to apply a time-dependent perturbation Hamiltonian $\Delta
{\cal H} (t) = \int V(x_0 , t) \; \rho (x_0) \; d^4 x_0$, where
$\rho(x_0) = \sum_{j=1}^N \delta^4(x_0 - x_j)$ is the density
operator; solve $i \hbar \, \partial_t | \psi (t) \rangle = [{\cal H}
  + \Delta {\cal H} (t) ] \; | \psi (t) \rangle$; and then compute the
expectation value $\expt{\psi (t) | \rho (x) | \psi (t)}$. Note that
the density operator here is four-dimensional, since the particles are
constrained to lie on a four-dimensional surface. To linear order in
$V (x , t)$ the result is completely specified by the density-density
response function
\begin{displaymath}
\expt{\rho (x) \rho(x_0)}_\omega = \! \! \! \!
\sum_{ \genfrac{}{}{0pt}{}
{\scriptstyle \expt{m|x_5|m} \le x_5^F}
{\scriptstyle \expt{m'|x_5|m'} > x_5^F}}
\biggl\{ \frac{\rho_{mm'}(x) \rho_{m' m}(x_0)}
{(\omega+i\eta) - (E_{m'} - E_{m})}
\end{displaymath}
\begin{equation}
+ \frac{\rho_{mm'}(x_0) \rho_{m'm}(x)}
{- (\omega+i\eta) - (E_{m'} - E_{m})} \biggr\}
\; \; \; ,
\label{pp}
\end{equation}
where
\begin{equation}
\rho_{m'm} (x) = \expt{ m' | \delta^4 (x) | m } /
\sqrt{ \expt{m' | m'} \; \expt{ m | m } } \; \; \; .
\end{equation}
The frequency $\omega$ refers to a particular Fourier component of the
time-dependent potential $V ( x , t)$. The ``infinitesimal'' $\eta$
corresponds physically to the inverse of the experiment duration.
Setting it to a small positive value slightly larger than the energy
level splitting, as we do here, amounts to the constraint that the
experiment never be done for a sufficiently long time to resolve this
splitting. Note that states involving multiple particle-hole
excitations do not contribute because their matrix elements vanish
identically.

We extract the momentum dependence of the response function \eqref{pp}
in the following way. We first set the number of fermions $N$ equal to
the number of Landau level states for which $\expt{x_5} \le 0$ (i.e.
half-filling), thus placing the edge at the equator $x_5 = 0$. The
wavefunctions and matrix elements are then easy to evaluate along the
great circle
    \begin{equation}
      x = (\sin \xi, \cos \xi, 0, 0, 0),
      \label{subspace}
    \end{equation}
which is contained in the edge. Note that the angle $\xi$ also gives
the distance along the great circle, since lengths are normalized to
the sphere radius. We find that
    \begin{multline}
      \rho_{m'm}(\xi) =  \frac{g_{m'm}}{p^2 \ell_0^4}
      \, (-1)^{m_3 + {m_3}'} \\
      \times e^{i\xi(m_3 - m_4 - {m_3}' + {m_4}')}
      \delta_{m_1 + m_4}^{m_1' + m_4'}
      \; \; \; , 
    \end{multline}
where
    \begin{equation}
      g_{m'm} \equiv \frac{(p+2)(p+3)}{2^{p+4} \pi}
                     \frac{(m_1 + m_4)!(m_2 + m_3)!}
                          {\sqrt{m_1! .. m_4! m_1'! .. m_4'!}}.
    \end{equation}
Note that $g_{m' m} = g_{m m'}$, as required by hermiticity of $\rho
(\xi)$. For numerical purposes, it is helpful to write \eqref{pp} as
the Fourier series
    \begin{equation}
      \expt{\rho(\xi) \rho (0)}_\omega = \frac{1}{p^4 \ell_0^8}
      \sum_{n = -p}^{p} f_n(\omega) e^{in\xi} \;,
      \label{series}
    \end{equation}
where the coefficients $f_n$ are given by
    \begin{equation}
      f_n(\omega) = \sum_{m,m',\pm}^{(n)} \;
      \frac{g_{m'm}^2}{\mp (\omega + i\eta) - (E_{m'} - E_m)}
    \end{equation}
with the sum taken over quantum numbers satisfying
    \begin{align}
    \begin{aligned}
      m_3 - m_4 - m_3' + m_4' &= \pm n \\
      m_1 + m_4 - m_1' - m_4' &= 0 \\
      m_1 + m_2 - m_3 - m_4 &\le 0 \\
      m_1' + m_2' - m_3' - m_4' &> 0 \\
      m_i, m_i' \ge 0\;, \quad
      \sum m_i &= \sum m_i' = p \; \; \; .
      \label{selection rules}
    \end{aligned}
    \end{align}
Note that this sum is invariant under the simultaneous interchange of
labels (1,2) and (3,4), so that $f_n (\omega) = f_{-n} (\omega)$. The
momentum-dependent response function is given in terms of these
quantities by
    \begin{align}
    \begin{aligned}
      &\expt{\rho_q \rho_{-q}}_\omega \\
        & = \frac{4 \pi (\sqrt{p} \, \ell_0 )^3}{Q}
           \int_0^\infty \expt{ \rho(\xi) \rho(0)}_\omega
           e^{- \alpha \xi^2} \sin(Q \xi) \; \xi d\xi \; \\
        & = \frac{(\sqrt{p} \, \ell_0 )^3}{Q}
           \left(\frac{\pi}{\alpha}\right)^{\frac{3}{2}}
           \sum_{n=-p}^p (Q + n) f_n(\omega) e^{-(Q + n)^2/4\alpha}
    \end{aligned}
    \label{fourier}
    \end{align}
where $Q \equiv q \sqrt{p} \, \ell_0$ and $\alpha$ is a convergence factor
required to remove standing-wave effects on the sphere.

    \begin{figure}
	    \includegraphics[scale=0.99]{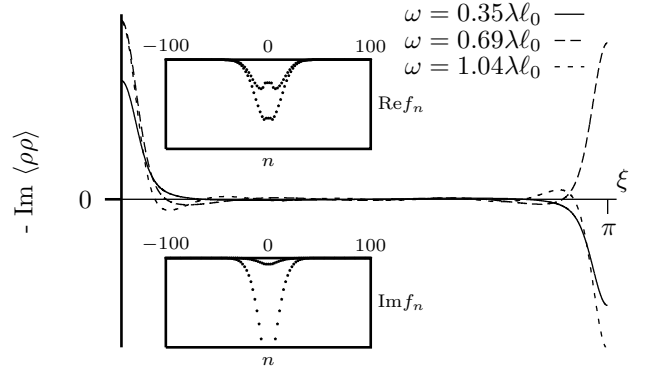}
      \caption{Plot of $- \mbox{Im}\, \expt{\rho(\xi) \rho
          (0)}_\omega$ against $\xi$ for different $\omega$, generated
        with $p = 300$ and $\eta = 0.087 \, \lambda \ell_0$. Inset:
        Fourier frequency components $f_n$ for $\omega = 0.35 \lambda
        \ell_0$. The two populations correspond to even and odd $n$,
        which are resonant at different frequencies.}
      \label{fig:space}
    \end{figure}

    \begin{figure}
	    \includegraphics[scale=0.99]{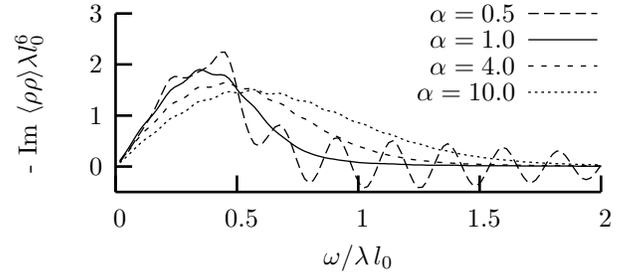}
      \caption{Plots of $- {\rm Im} \expt{\rho_q \rho_{-q}}_\omega$
        against $\omega$ for $\eta = 0.087 \, \lambda \ell_0$,
        $q\ell_0 = 0.25$, and $p = 300$.}
      \label{fig:alpha}
    \end{figure}

The need for a convergence factor is shown in Fig. \ref{fig:space},
where we plot the untransformed response function. There is a large
resonant response at the antipode $\xi = \pi$, which oscillates at the
frequency of the energy quantum $\Delta E = [2\sqrt{p}/(p+4)]
\lambda\ell_0 $. This antipodal response occurs because even and odd
$n$ correspond to different sets of energy levels in \eqref{selection
  rules}, and is an artifact of the finite system size. It disappears
in the thermodynamic limit, where the energy quantum goes to zero and
the sphere radius goes to infinity. A similar phenomenon occurs for
the 1D Fermi sea on a ring. In order to compare with the thermodynamic
limit, the desired convergence factor $\alpha$ is the smallest one
that removes the peak at the antipode, which otherwise produces a
long-range ringing in the Fourier spectrum. Larger values of $\alpha$
would relax momentum conservation unnecessarily. It may be seen from
Fig. \ref{fig:alpha} that the optimal value is about $\alpha \sim 1$.
We note in particular that the response at low frequencies is
diminished by increasing $\alpha$, indicating that it is physical and
not an artifact of $\alpha$.

The calculation has been tested in several important ways. For $p \leq
4$, which was sufficiently small to allow brute-force integration over
the spin angles $\theta$ and $\phi$, we verified the isotropy and
translational invariance of (\ref{pp}) and its accurate agreement with
(\ref{series}). The code generating the latter was then scaled up to
larger values of $p$ by changing a single parameter. We also evaluated
the Fourier transform in (\ref{fourier}) numerically and found it to
accurately match the analytic version.

Returning now to Fig. \ref{fig:pp vs q}, we observe that ${\rm Im}
\expt{\rho_q \rho_{-q}}_\omega$, as given by (\ref{fourier}), is
strikingly similar to the single-spin density-density response
function for the noninteracting Fermi sea in three spatial dimensions,
given by the well-known Lindhard function \cite{Pines}
    \begin{align}
    \begin{split}
     - & \mbox{Im} \, \expt{\rho_q \rho_{-q}}_\omega
      = \frac{q_F^3}{16\pi\omega_F} \\
      & \times \left\{ \begin{array}{l}
          \frac{\omega'}{q'}, \quad
            0 < \omega' < 2q' - {q'}^2 \;\;\mbox{and}\;\; q' < 2 \\ 
          \frac{1}{q'} \left[ 1 -
              \left(\frac{\omega' - {q'}^2}{2q'}\right)^2 \right],
          \; |2q' - {q'}^2| < \omega' < 2q' + {q'}^2 \\
          0 \quad \mbox{otherwise.}
        \end{array}\right.
    \end{split}  \nonumber \\
   & \qquad\qquad\qquad
    \omega' \equiv \omega / \omega_F \;,\;
    q' \equiv q / q_F
    \label{Lindhard}
    \end{align}
In either case one finds that ${\rm Im} \expt{\rho_q
  \rho_{-q}}_\omega$ is proportional to $\omega$ for small values of
$\omega$ and falls to zero at a threshold frequency $\omega = cq$. In
the case of the Fermi gas, $c$ is the Fermi velocity, or,
equivalently, the speed of compressional sound in the fluid. In the
case of the quantum Hall edge, $c \equiv 2 \lambda / p$ is the speed
of the ostensibly relativistic ``extremal dipoles'' found by Zhang and
Hu. These extremal dipoles are pair excitations which, for a given
energy, have the lowest possible momentum along the edge. The
non-extremal dipoles, which possess lower energies for each given
momentum, produce the response at $\omega < cq$. Elvang and Polchinski
\cite{Polchinski} have referred to the non-extremal edge states as
``tachyonic states''. However, this behavior is most simply
interpreted as that of a metal, in which case the extremal dipoles may
be identified by analogy as conventional sound.

We wish to thank B. A. Bernevig, J. P. Hu, D. I. Santiago, and S. C.
Zhang for encouragement and numerous helpful discussions. This work
was supported by the Department of Energy under Contract No.
DE-AC03-76SF00515. CYD acknowledges the Public Service Commission
(Singapore) for support.

\end{document}